# Longitudinal bunch monitoring at the Fermilab Tevatron and Main Injector synchrotrons[*]

R. Thurman-Keup,[†] C. Bhat, W. Blokland,[‡] J. Crisp,[§] N. Eddy, B. Fellenz, R. Flora, A. Hahn, S. Hansen, T. Kiper, A. Para, S. Pordes and A. V. Tollestrup

*Fermi National Accelerator Laboratory,*
*P.O. Box 500, Batavia, Illinois, USA*

E-mail: keup@fnal.gov

ABSTRACT: The measurement of the longitudinal behavior of the accelerated particle beams at Fermilab is crucial to the optimization and control of the beam and the maximizing of the integrated luminosity for the particle physics experiments. Longitudinal measurements in the Tevatron and Main Injector synchrotrons are based on the analysis of signals from resistive wall current monitors. This article describes the signal processing performed by a 2 GHz-bandwidth oscilloscope together with a computer running a LabVIEW program which calculates the longitudinal beam parameters.

---

[*] Work supported by Fermi Research Alliance, LLC under Contract No. DE-AC02-07CH11359 with the United States Department of Energy.
[†] Corresponding author.
[‡] Present address: Spallation Neutron Source, Oak Ridge National Laboratory, Oak Ridge, TN, USA.
[§] Present address: Facility for Rare Isotope Beams, Michigan State University, East Lansing, MI, USA.

# Contents

## 1. Introduction

Measurements of the behavior of a particle beam are crucial to the successful production and use of the beam, especially in environments where the beam must pass through several stages of accelerators. The state of an accelerated bunch of charged particles is determined by its six dimensional phase space volume which can be parameterized as a hyper-ellipse whose parameters are the covariances of the positions and momenta of all the particles relative to a particle following the ideal synchronous orbit (see section 5 of [1]). When motion in the three physical dimensions is uncorrelated, the beam can be characterized by the two dimensional phase space ellipse for each physical dimension. This paper discusses the determination of the longitudinal phase space, *i.e.* the dimension collinear with the beam, for Fermilab's Tevatron and Main Injector synchrotrons.

As stated above, the phase space variables are formulated as differences from the ideal orbit, which for the longitudinal phase space can be relative time of arrival at some point and momentum difference. A measurement of this longitudinal phase space ellipse requires measurements of the momentum spread, time distribution, and their correlation. However, since the aforementioned accelerators are circulating rings, one can make an assumption that, given enough time, the beam phase space density is stationary, and as such, can be parameterized by just one variable (see Liouville's Theorem for stationary states). The variable of choice from which to extract the phase space distribution is chosen to be the time distribution of the bunch. Note that in the Main Injector, the assumption of a stationary state is not entirely true, given its short cycle times. The details of how this is handled are discussed at the end of section 4.

The measurement of the time distribution of a bunch is typically made by a wall current monitor (WCM) which, by various means, samples the image currents induced on the wall of the beampipe [2-5]. The most recent incarnation at Fermilab is a resistive version [6,7] and is the basis of the measurements described in this article.

**2. Accelerator Environment**

The Main Injector and the Tevatron are the last two accelerators in a chain that starts with a 750 keV Cockcroft-Walton H$^-$ accelerator. The H$^-$ beam is fed to a linear accelerator where it is accelerated to 400 MeV before injection to the Booster Synchrotron where the beam is stripped of its electrons leaving just the protons. The Booster accelerates the beam to 8 GeV and sends it to the Main Injector. Antiprotons at Fermilab are produced from a target by 120 GeV protons from the Main Injector and are collected and cooled first by the Debuncher and Accumulator rings, and subsequently by the Recycler with an electron cooling system. There they await transfer to the Tevatron via the Main Injector. See [8] for a more thorough discussion of the Fermilab accelerator complex.

The Main Injector is a synchrotron with injection, ramping, and extraction happening on the time frame of 2 to 10 seconds depending on the cycle type. There are 4 major cycle types currently in use: proton injection to the Tevatron, antiproton injection to the Tevatron, antiproton transfers from Accumulator to Recycler, and a mixed cycle of antiproton production and/or delivery of protons to various fixed target experiments including the neutrino program. This variation in types of beams and time structures complicates the measurement of the longitudinal parameters.

The Tevatron is simpler in that it is *just* a storage ring. The only catch is that the protons and antiprotons are both present simultaneously in counter-circulating beams structured in 3 trains of 12 bunches each. Thus any monitor must be able to separate the various bunches in time.

Both the Tevatron and Main Injector have a ~53 MHz rf system resulting in rms bunch lengths that are a few nanoseconds. Additionally, the Main Injector has other rf systems for various manipulations including a 2.5 MHz system for handling antiprotons from the Accumulator or Recycler. These other rf frequencies imply that the measurement system must accommodate vastly different bunch lengths.

**3. Longitudinal Profile Monitor**

The longitudinal bunch monitors of the Tevatron and Main Injector are called the Sampled Bunch Displays (SBD). The original versions [9,10] used the same technique as the current version does: a high speed oscilloscope digitizes the signal from a WCM and a computer processes it [12]. The schematic of the current SBD is shown in figure 1. The source of the signal is a wide-band resistive WCM [6,7] installed in the F0(MI60) straight section of the Tevatron(Main Injector) which also houses the rf cavities. In the Tevatron, since protons and antiprotons circulate oppositely within the same beampipe, the WCM is located at F11 where the proton and antiproton bunches are maximally separated (~30 meters from the center of the straight section). The WCM has a >4 GHz bandwidth with a ~1 Ω gap resistance.

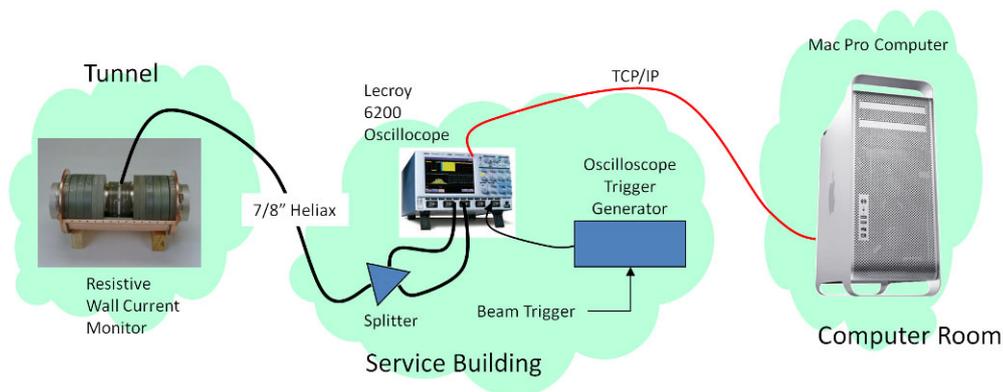

**Figure 1.** Schematic of the SBD. The splitter produces two identical signals which are captured by high-gain and low-gain channels in the oscilloscope. The oscilloscope trigger generator is the only major difference between the Tevatron and Main Injector versions. The Main Injector has a variety of operational cycles with different sampling requirements whereas the Tevatron version simply acquires data once a second continuously.

The WCM signal travels from the tunnel to an above ground service building through a 7/8 inch Heliax cable. There it is split and digitized by two separate channels of a Lecroy Corporation WaveRunner 6200 series oscilloscope having a 2 GHz bandwidth and 5 GSamples/s sampling rate. The two channels operate with unequal gains effectively increasing the dynamic range [11]. In the Tevatron, where the coexisting protons and antiprotons vary in intensity by more than a factor of three, the disparate gains allow simultaneous measurements of both particle species with equal sensitivity. In the Main Injector, the two gains allow for the variations in intensity experienced throughout, for example, coalescing cycles, where one starts with a number of low intensity bunches and ends with one high intensity bunch.

The oscilloscope trigger generation is controlled by a separate board. In the Tevatron case, the board is programmed to send 64 triggers to the oscilloscope, once every other revolution. The first 32 are triggered by the proton beam marker and the second 32 are triggered by the antiproton beam marker to obtain the relative antiproton timing. In the Main Injector, a specified number of triggers are sent at specified acquisition times within a cycle, which can last anywhere from 2 to 8 seconds. The acquisition times are chosen for their significance within the cycle; and the multiplicity of triggers at each acquisition enables averaging over synchrotron motion of the beam (see end of section 4).

After digitizing the total number of beam revolutions, the digitized data is sent to an Apple Mac Pro computer for signal processing and extraction of longitudinal beam parameters. All the processing is done within the National Instruments LabVIEW framework and the results are made available to the Accelerator Controls Network (ACNet).

### 3.1 Signal processing

Signal processing of the WCM data [12,13] consists of 3 main parts: combining of the 2 different gains to form one signal, correction of the dispersion in the cable from the WCM to the oscilloscope, and subtraction of the baseline.

### 3.1.1 Signal combining

Signal processing starts by combining the high-gain and low-gain signals obtained from the oscilloscope. Since the two signals follow separate electronic paths, they must first be adjusted for timing offsets before being combined. The offset between the two channels is determined by a convolution of the two channels and the high-gain signal is aligned to the low-gain using linear interpolation. Figure 2 shows a Main Injector high-gain/low-gain channel pair before combining. In combining the signals, high-gain channel samples that are close to the limit of the oscilloscope are replaced by their low-gain version (the line shown in figure 2).

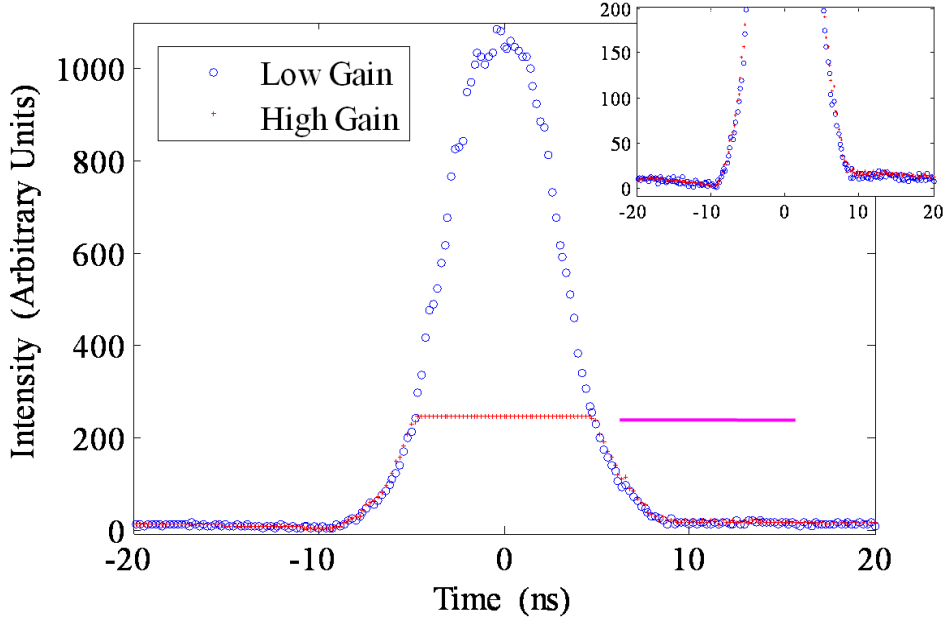

**Figure 2.** High and Low gain channels for a Main Injector antiproton bunch during injection to the Tevatron. The horizontal line shows the point above(below) which the combined signal is composed of the low-gain(high-gain) channel. The inset figure is an expanded view of the bottom section of the plot.

### 3.1.2 Cable dispersion

In the long cable path from the WCM to the oscilloscope, the signal spreads in time due to dispersion. The response of a coaxial cable to a step function, assuming only losses from conductor skin depths, is [14,15]

$$V_{out} = V_{in}\left(1 - \text{erf}\left[\frac{b\ell}{\sqrt{2t}}\right]\right) \quad (3.1)$$

where $b\ell$ is $1.45 \times 10^{-6}$ times the total attenuation in the cable at 1 GHz, $t$ is the time and erf is the error function[1]. The signal observed after the cable is a convolution of the input signal and the impulse response function of the cable which is just the derivative of equation 1. Truncating the impulse response to 40 ns and writing the convolution in discrete matrix form, one has

$$\mathbf{V}_{out} = \mathbf{G}\mathbf{V}_{in} \tag{3.2}$$

where **G** is a matrix with each row being the impulse response function shifted by one element from the previous row forming a triangular matrix. Inverting **G** leads to the solution for $\mathbf{V}_{in}$ with the last row of $\mathbf{G}^{-1}$ containing the desired finite impulse response (FIR) filter needed for correcting the dispersion. The FIR filter is convolved with $\mathbf{V}_{out}$ to obtain the original desired signal. Figure 3 shows the first few nanoseconds of the impulse response function and integral together with the FIR filter resulting from inverting **G**. The points are separated by 200 ps which is the sampling frequency of the oscilloscopes.

---

[1] $\text{erf}(x) = \int_0^x e^{-t^2} dt$

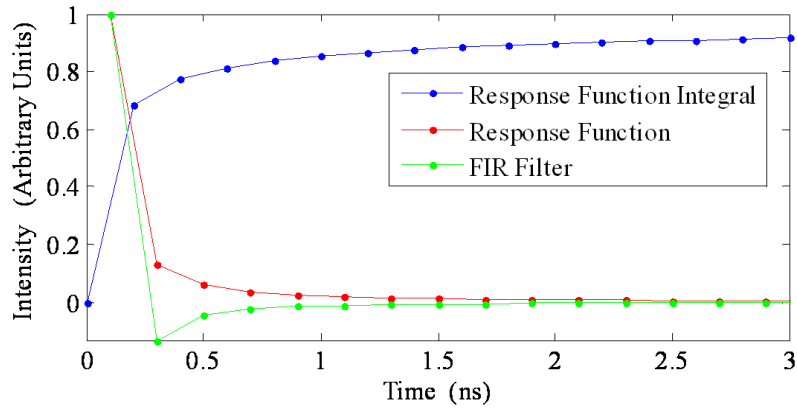

**Figure 3.** Plots of the first 3 ns of the response function and integral along with the FIR filter after inverting the response function matrix. The vertical axis units of the three functions are arbitrary and independent.

Figure 4 is a plot of the raw and dispersion corrected intensities for a Tevatron proton bunch with satellite bunches. The dispersion is obvious to the right of the main bunch and is completely removed after application of the FIR filter.

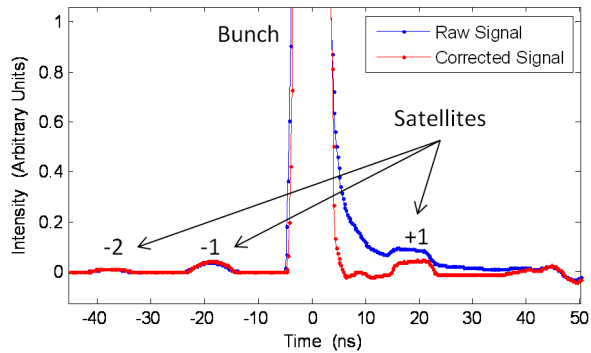

**Figure 4.** Plot of raw and corrected intensity signals for a Tevatron proton bunch and satellites. One can easily see the effect of removing the dispersion.

### 3.1.3 Baseline subtraction

The intensity data is divided into 18(5) sections in the Tevatron(Main Injector) and in each a histogram is made of the intensities. The baseline is taken to be the peak of the histogram. This technique requires the majority of points to be background points. Figure 5 shows the result of the baseline determination for a full revolution of data in the Tevatron.

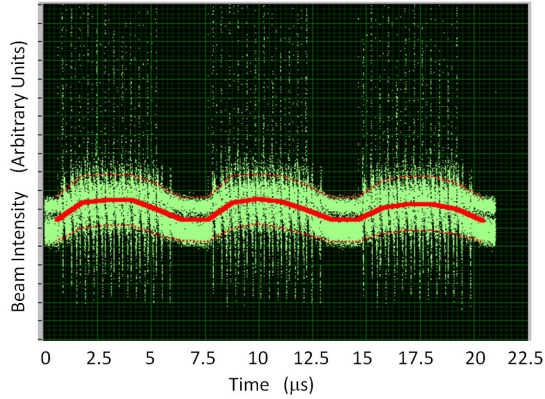

**Figure 5.** Plot of intensity data for a full revolution in the Tevatron. The green points are the data and the red curve is the baseline determined by histogramming the data points in each of 18 sections.

Figure 6 is a collection of aggregate bunches, meaning the individual bunches have been summed together, after adjusting for different centroids, to produce an average signal.

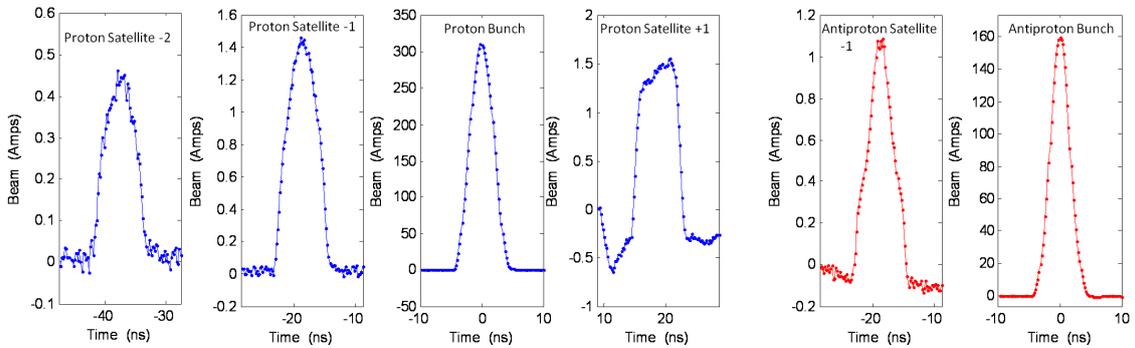

**Figure 6.** Intensity plots of Tevatron proton and antiproton bunches and several satellite bunches.

### 3.2 Longitudinal parameters

For each beam bunch, a variety of parameters are measured, including the intensity, the mean phase and rms width, the skew and kurtosis, and the emittance and momentum spread as calculated in section 4. An aggregate value (sum/average over all bunches) is calculated for each of the previous measurements and the whole collection is made available to users over ACNet.

### 4. Emittance analysis

The determination of the longitudinal emittance of the beams is accomplished by projecting phase space annuli onto the time axis and fitting these projections to the data [16] (see also [17,18] for other treatments of the subject). The beam is assumed to be in a stationary state and thus, in accordance with Liouville's Theorem, the phase space densities are constant in time and along phase space trajectories. The longitudinal Hamiltonian for this equilibrium beam is

$$\mathrm{H}(p,\varphi) = \Delta p^2 + \frac{2\beta^2 E_s e V_{rf}}{\eta h}\cos\left(\frac{2\pi}{\lambda_{rf}}z\right) \tag{4.1}$$

where $\Delta p$ is the momentum deviation from the synchronous particle, $E_s$ is the energy of the synchronous particle, $V_{rf}$ is the accelerating cavity voltage, $\eta$ is the slip factor, $h$ is the harmonic number (1113 for the Tevatron and 588 for the Main Injector), and $\lambda_{rf}$ is the wavelength of the accelerating voltage. Since H is constant along any trajectory, one can equate equation 4.1 to its value at $2\pi z/\lambda_{rf} = \pi$ (the center of the rf bucket) and solve for $\Delta p$

$$\Delta p = \pm\sqrt{\Delta p_0^2 - \frac{2\beta^2 E_s e V_{rf}}{\eta h}\left[\cos\left(\frac{2\pi}{\lambda_{rf}}z\right)+1\right]} \qquad (4.2)$$

where $\Delta p_0$ is the maximum momentum deviation on that trajectory. An example phase space with trajectories of constant density is shown in figure 7 with the projections of the annuli onto the z axis.

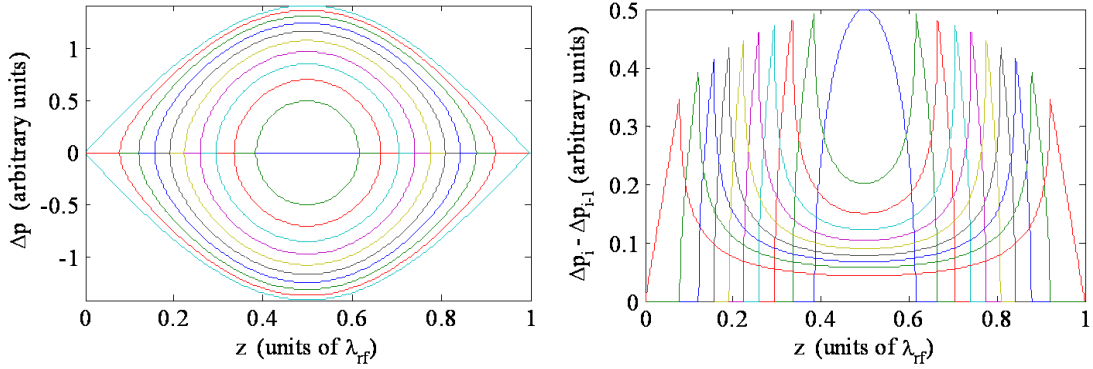

**Figure 7.** Left) Constant density trajectories in $\Delta p - z$ phase space. Right) Projections of the annuli onto the z axis. The projections are used as a set of basis functions in which the longitudinal beam distribution can be expanded. The annuli are chosen to have equal areas.

The longitudinal intensity distribution of the beam, $I(z)$, can be constructed from a properly weighted sum of the projections, $P_i(z)$,

$$I(z) = \sum_i w_i P_i(z) \qquad (4.3)$$

where $w_i$ is the relative particle density in annulus $i$, and is the quantity needed to determine the phase space distribution of the beam and hence the emittance. Since $I(z)$ is in reality a discrete sampling from the oscilloscope, equation 4.3 can be written as a matrix equation and inverted to find the weights,

$$\mathbf{I} = \mathbf{Pw}$$
$$\mathbf{w} = \mathbf{P}^{-1}\mathbf{I} \qquad (4.4)$$

The $\mathbf{P}^{-1}$ values for the Tevatron are calculated beforehand for both 150 GeV and 980 GeV beam and hard coded in the analysis program. For the Main Injector, the variety of operating modes necessitates recalculating the values of $\mathbf{P}^{-1}$ whenever the mode changes. To recalculate the values, the rf voltages and wavelengths and the beam energy, all of which are needed in equation 4.2, are obtained in real time for each measurement.

As discussed above, this technique requires an equilibrium beam. In the Tevatron, the beam circulates for many hours and is stable virtually the entire time. On the other hand, the Main Injector only contains beam for a few seconds at a time and never reaches an equilibrium condition. This manifests itself as asymmetries in the longitudinal distribution and is especially noticeable after coalescing where seven bunches are merged into a single bunch through rf manipulations and form higher density clusters in phase space. These clusters revolve in phase space at the synchrotron period leading to bumps in the longitudinal distribution that move in time (figure 8). To overcome this non-equilibrium beam, the longitudinal distribution is averaged over a synchrotron period and the average is fitted with the projections.

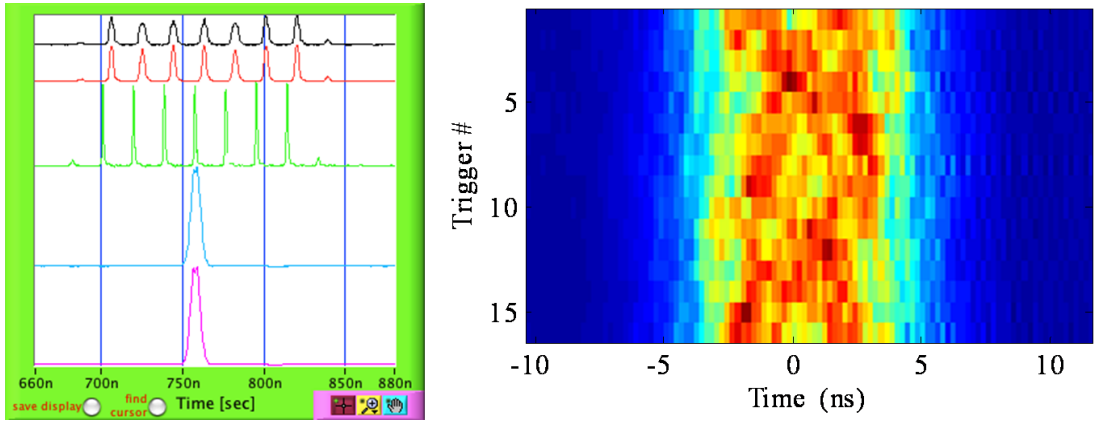

**Figure 8.** Left) Main Injector SBD data acquisition display. From top to bottom are successive Main Injector acquisitions for proton coalescing mode, where seven bunches (green trace) are merged into one large bunch (light blue trace). Right) Proton bunch time distributions for 16 triggers evenly spaced throughout the ~12 ms synchrotron period of the fourth acquisition in the left plot (light blue trace). These were collected immediately following coalescing of the seven proton bunches and demonstrate the synchrotron motion of the individual merged bunches.

The longitudinal emittance, which in the Tevatron and Main Injector is taken to be the phase space area containing 95% of the beam particles, is obtained by integrating equation 4.2.

$$\varepsilon_{95\%} = 2\int \sqrt{\Delta p_{95\%}^2 - \frac{2\beta^2 E_s eV_{rf}}{\eta h}\left[\cos\left(\frac{2\pi}{\lambda_{rf}}z\right)+1\right]}\,dz \qquad (4.3)$$

Here $\Delta p_{95\%}$ is the peak of the contour within which 95% of the beam is contained. This contour is determined from the weights returned by the above fitting procedure. Since the phase space variables are $z$ and $\Delta p_z$, which are canonical variables, the emittance can be considered a normalized emittance and should not change with beam acceleration.

Figure 9 shows the growth of the Tevatron longitudinal emittance over the course of a store. This emittance growth is expected and has been discussed in the literature [19,20]. As a test of the accuracy of the measurement technique, figure 10 shows two Tevatron comparisons: the longitudinal parameters before and after acceleration, and a comparison with the 1.7 GHz Schottky detector [21].

At the time this emittance measurement technique was developed, a significant effort was underway to investigate the low luminosity in the Tevatron. The longitudinal emittances were calculated by assuming that the beam phase space was separable in $\Delta\phi$ and $\Delta E$, a more stringent constraint than uncorrelated, leading to Gaussian distributions in the two variables. The consequence of assuming Gaussian distributions in the case of the Tevatron was a large calculated emittance growth during acceleration (>20%), contrary to expectations. Since there is no a priori reason to expect the Tevatron beam longitudinal variables to be separable, the emittance technique in this paper was developed. It demonstrated that there was in fact no significant growth in the emittance [22]. This outcome prevented much effort hunting down non-existent problems in the rf acceleration systems. The left plot of figure 10 shows the emittance at 150 GeV, before acceleration, and the emittance after acceleration to 980 GeV. As can be seen the emittance after acceleration is essentially unchanged from before.

The comparison with the Schottky detector is important in that it is an independent device. There is no arbitrary scaling involved in the right plot in figure 10 and since the SBD measurement of the momentum spread is a direct by-product of the emittance algorithm, the comparison is a direct test of the technique.

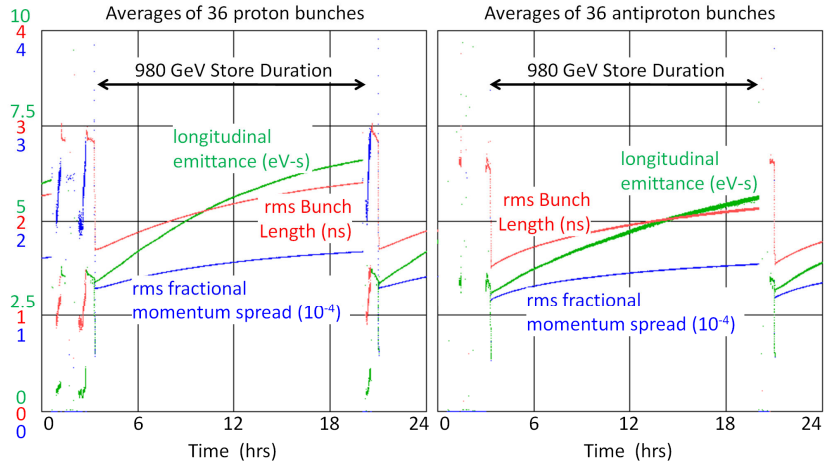

**Figure 9.** The growth of the measured longitudinal emittance, momentum spread and bunch length over the course of a single store in the Tevatron. The left plot is for the protons and the right plot is antiprotons. The arrows indicate the time during which the Tevatron beam was at 980 GeV undergoing collisions.

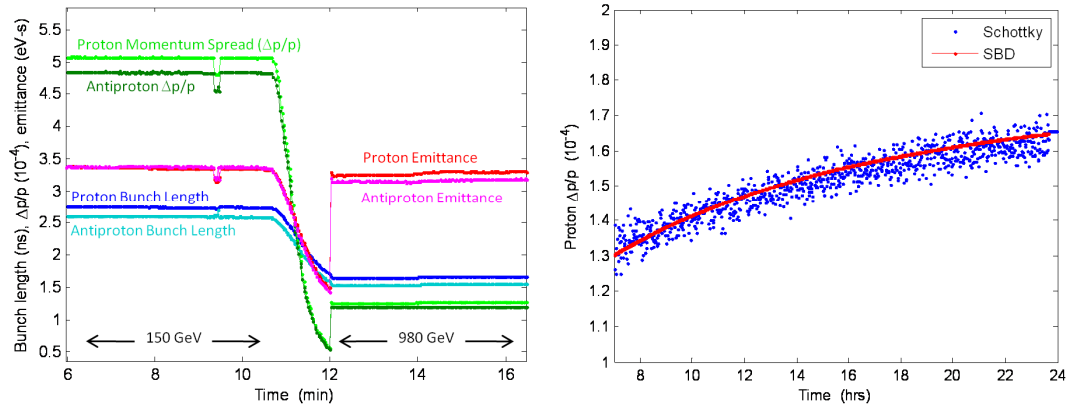

**Figure 10.** Left) Comparison of longitudinal parameters before and after acceleration from 150 GeV to 980 GeV. The bunch length changes as does the momentum spread ($\Delta p/p$), but the emittance is nearly the same as expected. Right) Comparison of the momentum spread as measured with the SBD using the emittance algorithm, with that obtained from the 1.7 GHz schottky detector. The measurements are absolute (i.e. no arbitrary scaling between the devices) and show excellent agreement.

Figure 11 shows a comparison of longitudinal emittances as measured in the Recycler and in the Main Injector. These measurements are of antiprotons as they are transferred from the Recycler to the Main Injector on their way to the Tevatron, and are made while the bunch is very long (50-100 ns) before being split up and coalesced. The agreement is quite good.

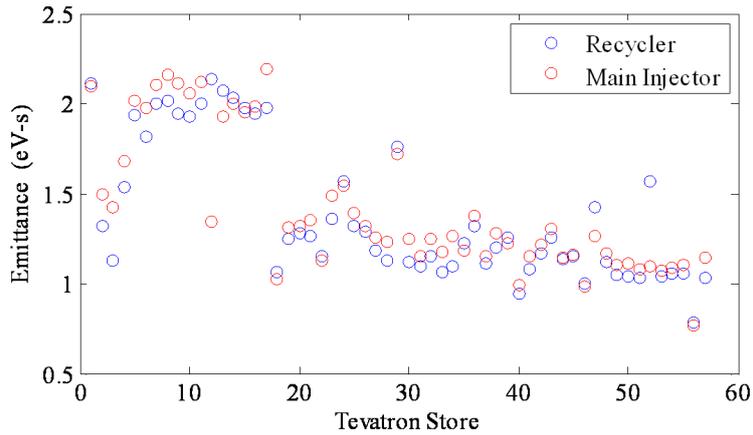

**Figure 11.** Comparison of longitudinal emittance measured in the Recycler and the Main Injector for nearly 60 transfers of antiprotons destined for the Tevatron. The Main Injector measurements are taken when the bunch is still very long (the rf is running at 2.5 MHz).

## 5. Operational use

The SBD is used as a regular part of operations to control the quality of the beam that arrives in the Tevatron. The longitudinal emittance of antiprotons arriving in the Main Injector from the Recycler is required to fall within a certain window. The bunch length of protons in the Tevatron are similarly constrained, and if these limits are exceeded, either adjustments must be made or the beam is discarded.

In addition to everyday operational use, the SBD is used for less frequent adjustments. The bunch shape from the SBD is used to resolve timing issues in the Main Injector that occur anywhere from before injection to after coalescing. The Main Injector beam loading compensation is tuned to equalize the emittance of each bunch after coalescing. The Booster extraction kicker timing is adjusted to extract exactly 7 bunches using the SBD information. In the Tevatron, it has been used to study longitudinal instabilities as shown in figure 12 (also see [23,24]). The fine time resolution allows one to observe beam loading of the rf cavities in the Tevatron (figure 13).

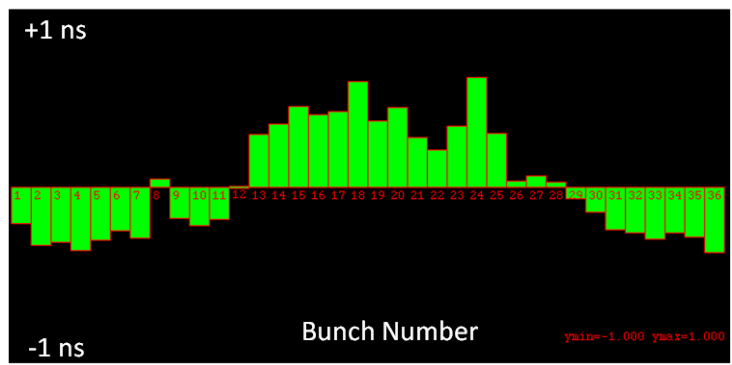

**Figure 12.** Histogram of Tevatron bunch centroids during a longitudinal mode 1 instability.

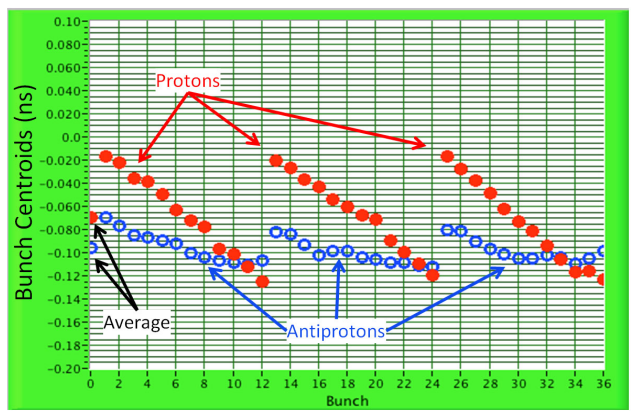

**Figure 13.** Plot of centroids of proton and antiproton bunches. The time resolution of the SBD is clearly at the level of a few picoseconds and allows one to see a clear slope in the proton bunch trains which is believed to be due to beam loading of the rf cavities.

## 6. Conclusions

The SBDs have functioned well throughout Tevatron Run II allowing for precise fine tuning of the longitudinal dynamics of the beam. The increased dynamic range allowed for the measurement of the satellite bunch intensities in the Tevatron. The fine time resolution results in detailed measurements of bunch-to-bunch variations and, together with the phase space fitting algorithm, precise longitudinal emittances.

## Acknowledgments

The authors thank their colleagues throughout the Accelerator Division for their assistance and co-operation in making the SBD a robust and precise measurement device.